\documentclass[twocolumn]{aastex631}
\makeatletter
\let\frontmatter@title@above=\relax

\usepackage{todonotes}
\usepackage{amsmath}
\usepackage{graphicx}
\usepackage{csquotes}

\begin{document}

\title{Prospects for identifying pulsar candidates in radio surveys using scintillation}

\author{Jitendra Salal}
\affiliation{National Centre for Radio Astrophysics, Post Bag 3, Ganeshkhind, Pune, 411007, India} 

\author{Shriharsh P. Tendulkar}
\affiliation{CIFAR Azrieli Global Scholars program, CIFAR, Toronto, Canada}

\author{Visweshwar Ram Marthi}
\affiliation{National Centre for Radio Astrophysics, Post Bag 3, Ganeshkhind, Pune, 411007, India}

\begin{abstract}
In our previous paper, we developed a technique for identifying pulsar candidates in interferometric radio images using their distinctive scintillation signatures. Building on this technique, the present study simulates a pulsar population using the \texttt{PsrPopPy} Python module to investigate the technique's limitations and detection capabilities. Among pulsars detectable exclusively by this technique, 50\% have duty cycles exceeding the mean value of 0.09 observed in time-domain detections. Our pulsar population simulations revealed a set of observational parameters that optimize pulsar detection. An observation frequency of $~$1420\,MHz and a channel width of $~$10\,kHz emerge as the optimal configuration to maximize the pulsar detection efficiency. By applying a scintillation-based technique to future radio telescopes like DSA-2000, we can detect 56\% of normal pulsars and 84\% of MSPs in addition to those detected using non-imaging, time-domain surveys. These detected pulsars cannot be verified by time-domain searches.

\end{abstract}

\section{Introduction} \label{sec:intro}
Time-domain search techniques, which identify pulsars through their periodic emission, are successful for the detection of many pulsars, but struggle to detect several important subpopulations that exhibit high-duty-cycle pulsars with broad pulse profiles, binary systems in which orbital motion obscures the underlying periodicity, and nearly aligned rotators with small angles between their magnetic and rotation axes \citep{young}. Such selection effects introduce substantial biases in our understanding of the Galactic neutron star population.\\

Existing radio telescopes, including the upgraded Giant Metrewave Radio Telescope \citep[uGMRT;][]{yashwant} and the Green Bank Telescope \citep[GBT;][]{gbt2022}, have stored extensive archival datasets spanning decades of observations. Collectively, radio facilities worldwide have discovered thousands of pulsars; a significant population remains undetected as a result of limited sensitivity to faint pulsars and inherent limitations in time-domain searches. The challenges arise from interstellar effects such as dispersion smear pulses across frequency channels, while scattering broadens pulse profiles, both reducing the effectiveness of time-domain searches.\\

The next generation of radio surveys, such as those planned with the Square Kilometre Array (SKA) \citep{pradoni2015} and precursor instruments such as MeerKAT and the Australian SKA Pathfinder (ASKAP) \citep{norris2011}, will address some of the limitations. These facilities employ wide bandwidths, high time resolution capabilities, and interferometric baselines that increase sensitivity and angular resolution. The resulting surveys will generate ultra-deep, wide-field images of the radio sky \citep{norris2011}, cataloging millions of compact radio sources with unprecedented precision. Within this population, radio pulsars will constitute a small fraction of the order of 10$^{-5}$ \citep{keane2015}, but remain critical for probing fundamental physics, including neutron star equation-of-state constraints \citep{Demorest2010}, tests of general relativity in strong gravitational fields \citep{stairs2003}, and the detection of low-frequency gravitational waves through pulsar timing arrays \citep{Agazie2023}.\\

The Five-hundred-metre Aperture Spherical Radio Telescope (FAST), with its exceptional sensitivity, is expanding the pulsar population by detecting sources too faint to be detected by previous generations of radio telescopes \citep{fast}. However, certain pulsar populations remain challenging to detect using time-domain searches. These include high-duty-cycle pulsars, which produce broader pulse profiles that reduce the effectiveness of period folding techniques, and pulsars in binary systems, where orbital motion complicates the time-domain search.\\

Previous efforts have demonstrated the potential of image‑domain searches for pulsars. Early work showed that pulsars often appear as compact, steep‑spectrum radio sources \citep{backer}. This approach was subsequently developed in a number of spectral studies of wide‑field surveys to identify pulsar candidates \citep{Kaplan_2000}, while other efforts used continuum imaging to recover known pulsars and search for new ones \citep{frail2018}. More recently, wide‑field continuum studies have highlighted that pulsars stand out as highly linearly and circularly polarized sources \citep{sobey2022, Wang_2022}. Similar image‑based searches with the Murchison Widefield Array (MWA) have also used steep spectra, polarization, and variability to identify pulsar candidates \citep{sett2023}.\\

Although these approaches have proven effective, they remain limited by contamination from other compact sources with steep spectra or high polarization fractions, such as AGN and some Galactic radio stars. Our work builds on pulsar time-variability based on diffractive interstellar scintillation, we have developed a complementary detection method that identifies pulsar candidates through their scintillation signatures in interferometric data and then confirms them via time-domain searches. Our scintillation-based visibility correlation searches (SVCS) technique cross-correlates dynamic spectra derived from visibility data to measure scintillation bandwidths and timescales as detailed in our previous work \citep[hereafter S24]{salal}.\\

This paper quantifies the SVCS capabilities through numerical simulations in diverse pulsar populations. We evaluated the number of pulsar candidates we can discover in radio images under various observational conditions and instrumental configurations. The paper is structured as follows. Section \ref{sec:simulpulsar} presents our pulsar population synthesis methodology, detailing the statistical distributions and physical assumptions underlying our simulated pulsar populations. Section~\ref{sec:Detectionpulsar} describes the detection algorithms and criteria used in different search techniques, including time-domain searches, imaging analysis, and our scintillation-based approach. Sections \ref{sec:gmrtsurvey} and \ref{sec:radiosurvey} demonstrate practical applications of our population synthesis framework to the GMRT and contemporary radio continuum surveys. Section \ref{sec:analysis} examines the implications of our simulation results, analyzing detection efficiencies and limitations across different pulsar parameters. Finally, Section \ref{sec:conclusion} summarizes our findings and discusses future applications of this technique for upcoming radio surveys.\\

\section{Simulation of Pulsars}
\label{sec:simulpulsar}
We used the \texttt{PsrPopPy} Python package \citep{psrpoppy} to generate statistically representative simulations of the pulsar population. Our study extends the package's core functionality by implementing new modules that simulate millisecond pulsar populations, pulsar detection in radio images, and SVCS detection as described in S24 through SNR determination. The \texttt{PsrPopPy} framework employs a snapshot generation approach that requires initial calibration against existing pulsar surveys. The number of pulsars detected in a reference survey serves as input and provides a statistical anchor for the population synthesis. The package generates simulated pulsar populations by sampling from a series of statistical models that describe key pulsar parameters: pulse period, period derivative, luminosity, Galactic spatial distribution, spectral index, and duty cycle.\\

The population generation algorithm continues iteratively until the specified number of detectable pulsars matches the reference survey's detection statistics. This approach accounts for the complex observational selection effects that typically exclude certain pulsar populations from detection, such as faint pulsars below instrumental sensitivity thresholds, smeared sources with reduced coherence, pulsars located outside the survey region, and sources with unfavorable geometric configurations.\\

Our current simulation framework focuses on two primary pulsar populations: normal and millisecond pulsars (MSPs). Although many fundamental parameter distributions remain consistent between both populations, we incorporate population-specific variations where empirical evidence suggests significant differences. The subsequent sections will detail these population-specific parameter distributions.\\

\textbf{Modeling Interstellar Scintillation Parameters with Galactic Electron-Density Distributions:}\\
The three-dimensional Galactic electron density distribution model YMW16 \citep{Yao2017} generated estimates of dispersion measure (DM) for pulsars at specified Galactic longitudes, latitudes, and distances. This model resolved large-scale structures in the Milky Way ionized interstellar medium, including spiral arms and local electron density fluctuations.\\

\textbf{Radial distribution:} We used a radial distribution model for normal pulsars and MSPs, derived from the Galactic population study by \cite{lorimer2006}. The model describes the surface density of pulsars as a function of the galactocentric radius, capturing the spatial structure of pulsar populations within the Milky Way.

\begin{equation}
    \rho(r)\propto\left(\frac{R}{R_{\odot}}\right)^{a}\mathrm{exp}\left[-b\frac{R-R_{\odot}}{R_{\odot}}\right]
\end{equation}

Here, $\rho(R)$ represents the surface density at galactocentric radius $R$; $R_{\odot}$ denotes the Sun's distance from the Galactic centre; $a = 1.9$ and $b = 5$ are model parameters.\\

\textbf{Galactic z coordinate distribution:}  In Galactic z coordinates, the distribution of normal pulsars and millisecond pulsars was approximated by a two-sided exponential

\begin{equation}
    N(z)\propto \mathrm{exp}\left(\frac{-|z|}{E}\right)
\end{equation}

Here, E is the scale height. For normal pulsars E = 0.33\,kpc \citep{lorimer2006} and for MSPs E = 0.5\,kpc, as suggested by \cite{levin2013}.\\

\textbf{Spectral index distribution:}
We adopted a Gaussian distribution of spectral indices for normal pulsars and MSPs to characterize the radio emission properties. We took a mean spectral index of -1.6 and a standard deviation of 0.35 \citep{lorimer1995} for normal pulsars and MSPs.\\

\subsection{Normal pulsars}

\textbf{Pulse period distribution:} We used a log-normal distribution model to characterize the pulse period characteristics of normal pulsars \citep{lorimer2006}. This distribution incorporates the beaming model developed by \cite{smits2009}.

\begin{equation}
    f(P)\propto\mathrm{exp}\left[-\frac{(log\,P-\mu)^2}{2\sigma^2}\right]
\end{equation}

where $\mu$ = 2.7 is the mean and $\sigma$ = $-0.34$ is the standard deviation.\\

\textbf{Luminosity distribution:} To characterize normal pulsar luminosities, we took a log-normal probability distribution with a mean of log$_{10}$L = -1.1 and a standard deviation of log$_{10}$L = 0.9 \citep{faucher}.\\

\subsection{Millisecond Pulsars}

\textbf{Pulse period distribution:} We adopted a log-normal distribution model to characterize the pulse period of MSPs \citep{lorimer2015}.

\begin{equation}
    f(P)\propto \frac{1}{P}\mathrm{exp}\left[-\frac{(lnP-\mu)^2}{2\sigma^2}\right]
\end{equation}

where $\mu$ = 1.5 and $\sigma$ = 0.58.\\

\textbf{Luminosity distribution:} We used a power-law luminosity distribution for MSPs \citep{levin2013}.

\begin{equation}
    N\propto\left(\frac{L}{\mathrm{mJy\, kpc^2}}\right)^{-1.45}
\end{equation}

with L (luminosity) ranging from 0.1\,mJy\,kpc$^2$ to 100\,mJy\,kpc$^2$.

\subsection{Binary Pulsars}

Binary pulsars, defined as systems in which a pulsar is gravitationally bound to a companion star, present unique detection challenges due to orbital motion. The detection sensitivity for binary pulsars is reduced when time-domain search methods are applied, as orbital acceleration and higher-order kinematic terms (jerk) introduce phase variations in the pulsar signal. To mitigate this loss, acceleration and jerk search algorithms have been implemented in pulsar surveys. Although these methods recover most of the lost sensitivity, the large parameter space associated with orbital motion requires significant computational time to search.\\

Although binary pulsar simulations were not conducted in this study, the detection challenges for such systems resemble those encountered in normal and millisecond pulsar searches. However, binary pulsars present additional complexity in time-domain searches because of their reduced detection sensitivity.\\

\section{Detection of Pulsars in searches}
\label{sec:Detectionpulsar}

\subsection{Detection in radio images}
In radio images, pulsars can be identified as compact, steep-spectrum \citep{backer}, highly circularly polarized \citep{Wang_2022}, and highly linearly polarized sources \citep{sobey2022}. In the pulsar population simulation, we predict the number of pulsars detected in the radio image by setting the cutoff for image SNR\,$>$\,5. The image SNR is defined as
\begin{equation}
\label{eqn:image}
    SNR_I=\frac{S}{\sigma}.
\end{equation}

Here, S is the flux density and $\sigma$ is the system noise introduced by various factors during the observations and can be calculated using the radiometer equation.
\begin{equation}
    \sigma=\frac{T_{s}}{G\sqrt{N_a(N_a-1)BWT_{obs}}}.
\end{equation}

Here, $T_s$ is the system temperature, G is the antenna gain, $N_a$ is the number of antennas, BW is the observed bandwidth, and $T_{obs}$ is the total observed time.\\

\subsection{Detection in time-domain searches}
In a time-series analysis, a pulsar appears as a periodic signal after de-dispersion. The \texttt{PsrPopPy} package defines the SNR for time-domain searches as:

\begin{equation}
\label{eqn:pulsar_survey}
\text{SNR}_p = \frac{S}{\sigma}\sqrt{\frac{2}{N_a-1}}\sqrt{\frac{1-\delta}{\delta}},
\end{equation}

where $\delta$ represents the duty cycle of the pulsar, calculated by:

\begin{equation}
\delta = W_{\text{eff}}/P.
\end{equation}

Here, $W_{eff}$ denotes the effective pulse width, P is the pulsar period, and the default periodic SNR cutoff is 9. The effective pulse width is determined by:

\begin{equation}
W_{\text{eff}} = \sqrt{W^2_{\text{int}} + t^2_{\text{samp}} + \Delta t^2 + \tau^2_{\text{sc}}},
\end{equation}

where $W_{int}$ is the intrinsic pulse width, $t_{samp}$ is the sampling time, and $\Delta t$ represents the dispersive smearing time:\\

\begin{equation}
\Delta t = 8.3 \times 10^6 \times \text{DM} \times \frac{\Delta\nu_{\text{MHz}}}{\nu^3_{\text{MHz}}} \text{ms}.
\end{equation}

The pulse smearing due to interstellar scattering, $\tau_{sc}$ , follows the relation from \citep{krishnakumar2015}:
\begin{equation}
\tau_{\text{sc}} = 4.1 \times 10^{-11}\, \text{DM}^{2.2}(1.0 + 0.00194\, \text{DM}^2),
\end{equation}

\subsection{Detection via SVCS}
In the cross-correlation of dynamic spectra as discussed in S24, pulsars can be identified as bright elliptical structures at the centre formed by the scintillation of the pulsar signal. We used the correlation SNR with a cutoff of 5 to detect pulsars.

\begin{equation}
\label{eqn:correlation}
    SNR_{cor}=\frac{S^2m^2}{\sigma^2}\sqrt{\frac{\pi\tau_s f_s}{2T_{obs}BW}}.
\end{equation}

Here, m is the modulation index, $\tau_{s}$ and f$_{s}$ are the scintillation timescale and frequency.\\

\textbf{Modulation index:} The modulation index for diffractive scintillation is 1. The observed modulation index is related to the time and frequency resolution of the telescope \citep{handbook}.
\begin{equation}
    m_{obs}=\frac{m}{\sqrt{N_tN_f}}=\frac{1}{\sqrt{N_tN_f}}.
\end{equation}
\begin{equation}
\label{eqn:kappa}
    N_t=1+\kappa\frac{\Delta t}{\tau_s},
    N_f=1+\kappa\frac{\Delta\nu}{f_s}
\end{equation}

Here, $\kappa$ is 0.15. $\Delta t$ and $\Delta\nu$ are the sub-integration time and channel width, respectively.\\

\textbf{Scintillation parameters:} The scintillation parameters can be calculated using the formula \citep{cordes1991}
\begin{equation}
\label{eqn:stime}
    \tau_s=3.3\, \nu_{GHz}^{6/5}SM^{-3/5}V^{-1}_{100}\,s.
\end{equation}

\begin{equation}
\label{eqn:sfreq}
    f_s=223\, \nu_{GHz}^{22/5}SM^{-6/5}D^{-1}_{kpc}\,Hz.
\end{equation}
We are using the NE2001 model \citep{ne2001} to calculate the scintillation parameters because the YMW16 model does not have information on the scattering measure (SM). For pulsars, the typical scintillation timescale is roughly minutes, and the scintillation bandwidth is roughly MHz at $\sim$1\,GHz \citep{rickett1990}.\\

\section{GMRT survey}
\label{sec:gmrtsurvey}
\begin{deluxetable*}{lcccccccc}
\tabletypesize{\scriptsize}
\tablewidth{0pt} 
\tablecaption{Simulated detection statistics for normal pulsar population observed with the GMRT survey under varying observational parameters\tablenotemark{a}. The table compares pulsars identified in radio image, time-domain and SVCS across three GMRT frequency bands, with variations in bandwidth and spectral resolution.
\label{tab:normal}}
\tablehead{
\colhead{Frequency} & \colhead{Bandwidth} & \colhead{Channels} & \colhead{Time-domain} & \colhead{Smeared} & \colhead{Image} & \colhead{SVCS} & \colhead{Common} & \colhead{Pulsar candidates}\\
\colhead{MHz} & \colhead{MHz}
} 
\startdata
325& 16& 128 & 1505& 60368 & 2366 & 527& 505 & 21\\
610& 16& 128 & 1072& 43945& 1662& 593 & 506 & 61\\
"& 200& 2048 & 2784& "& "& 890& 841& 43\\
"& "& 4096 & 2791& "& "& 889& 841& 42\\
1420& 16& 128 & 370& 27039& 591& 467& 275& 102\\
"& "& 256 & 383& "& "& 467& 274& 103\\
"& "& 512 & 387& "& "& 468& 281& 97\\
"& 32& 256 & 528& "& 820& 547& 354& 119\\
"& "& 512 & 532& "& "& 548& 356& 120\\
"& 200& 2048 & 1224& "& 1843& 829& 667& 103\\
"& "& 4096 & 1236& "& "& 834& 676& 99\\
"& "& 8192 & 1245& "& "& 835& 631& 145\\
"& "& 16384 & 1251& "& "& 835& 677&  99\\
\enddata
\tablenotetext{a}{Pulsar population was simulated by using the Parkes multibeam survey detections (N=1038) as fiducial model to produce 120,000 simulated pulsars. In time-domain searches for the GMRT survey, a sampling time of 0.1\,ms is assumed. The results are presented in columns showing: (4) pulsars detected in time-domain searches; (5) pulsars lost due to dispersion smearing; (6) detections in radio images; (7) detections in SVCS; (8) pulsars detected by all three methods (radio imaging, time-domain searches, and SVCS); and (9) pulsars detected exclusively through SVCS.}
\end{deluxetable*}

The GMRT represents a radio interferometric array of 30 steerable parabolic antennas, each measuring 45 meters in diameter. These antennas are distributed over a region that spans 25 km in diameter, following an approximately Y-shaped configuration. The array combines 14 antennas within a compact central array of 1 square kilometre and 16 antennas along three outward extending arms, enabling high sensitivity and excellent spatial resolution.\\

We conducted pulsar population simulations in multiple GMRT frequency bands at 325\,MHz (band\,3), 610\,MHz (band\,4), and 1420\,MHz (band\,5) to evaluate the detection capabilities of our scintillation-based identification technique. These frequency bands offer complementary advantages: lower frequencies provide higher sensitivity to steep-spectrum sources, while higher frequencies minimize dispersion and scattering effects at the same channel width.\\

Our simulations incorporated varying spectral resolutions, with the total frequency channels ranging from 128 to 16,384. This wide range of channelization enables a detailed investigation of how frequency resolution affects scintillation detection efficiency. The channel configurations correspond to frequency resolutions spanning from coarse (a few hundred kHz) to fine (tens of kHz) sampling, allowing us to optimize detection sensitivity across different DM regimes.\\

We assumed 26 operational antennas for these simulations, reflecting typical GMRT observing conditions in which some antennas may be unavailable due to maintenance or technical considerations. This configuration still provides sufficient baseline coverage and sensitivity for robust interferometric imaging and scintillation detection.\\

The array's frequency flexibility, wide field of view, and significant collecting area make the GMRT particularly well suited to large-scale pulsar surveys using our scintillation-based detection methodology. The interferometric nature of the telescope enables simultaneous imaging and time-domain analysis \citep{yashwant}, maximizing the scientific return of each observation.\\

\subsection{Normal Pulsars}

We conducted a simulation of the pulsar population using the \texttt{PsrPopPy} to generate a statistically representative sample of Galactic normal pulsars. The simulation was normalized using the Parkes Multibeam Pulsar Surveys \citep{manchester2001} as a reference, matching the observed count of 1038 detected pulsars. This normalization yields an underlying Galactic population of 120,000 normal pulsars (including those beamed away from Earth), consistent with the predictions of \citep{faucher}. Adopting the PMSURV survey's normal pulsar population distribution as our fiducial model, we simulated GMRT surveys by varying observational parameters. From this simulated population, we identified pulsars detectable in radio images, time-domain searches, and SVCS when their signals exceeded the respective detection thresholds described in Section~\ref{sec:Detectionpulsar}.\\

Table~\ref{tab:normal} presents a \texttt{PsrPopPy} simulation of the normal pulsars in three frequency bands with different observational bandwidths and frequency resolutions with an observation time of 30\,min and a time resolution of 2\,s in interferometric radio data and 128\,$\mu$s in time-domain data. Comparing the bands with 16\,MHz bandwidth and 128 channels, we found that the highest number of pulsar candidates uniquely identified through SVCS occurs at 1420\,MHz. At this frequency, increasing the observational bandwidth increases detections in both SVCS and time-domain searches, although the number of candidates detectable only through SVCS remains approximately constant across bandwidths, peaking at 200\,MHz bandwidth with 8,192 channels.\\

The central frequency of the 610\,MHz band with the 200\,MHz bandwidth yields the highest overall detection rate in SVCS. This optimal performance results from two competing effects: Pulsars exhibit steep spectra and are brightest at lower frequencies, where their flux density decreases significantly with increasing frequency, while according to Equation~\ref{eqn:sfreq}, scintillation parameters improve with frequency. Equation~\ref{eqn:correlation} shows that the correlation SNR consequently increases with frequency. At 610\,MHz, these opposing effects balance to produce the maximum theoretical SVCS detection rate. Although many pulsars at this frequency remain detectable through time-domain searches, SVCS identifies a significantly larger number of pulsars missed by time-domain searches, owing to its superior detection efficiency at this particular frequency.\\

We selected a central frequency of 600\,MHz with the 200\,MHz bandwidth as our reference configuration for pulsar population simulations. Although this frequency suffers from substantial RFI, 1420\,MHz provides comparable detection rates and the highest number of uniquely identifiable pulsar candidates through SVCS against practical observational constraints.\\

\begin{figure}
\includegraphics[width=0.5\textwidth]{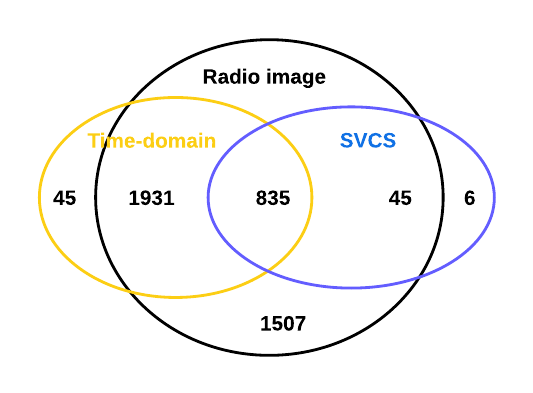}

\caption{Venn diagram for simulated pulsars detected by three techniques. The black, yellow, and blue circles corresponds to normal pulsars detected in radio images, pulsar surveys, and SVCS. Pulsar population simulation is conducted at a frequency 600\,MHz with bandwidth 200\,MHz, observation time 30\,minutes and sub-integration time 2\,s.}
\label{fig:normpuls}
\end{figure}

\begin{figure*}
\includegraphics[width=0.5\textwidth]{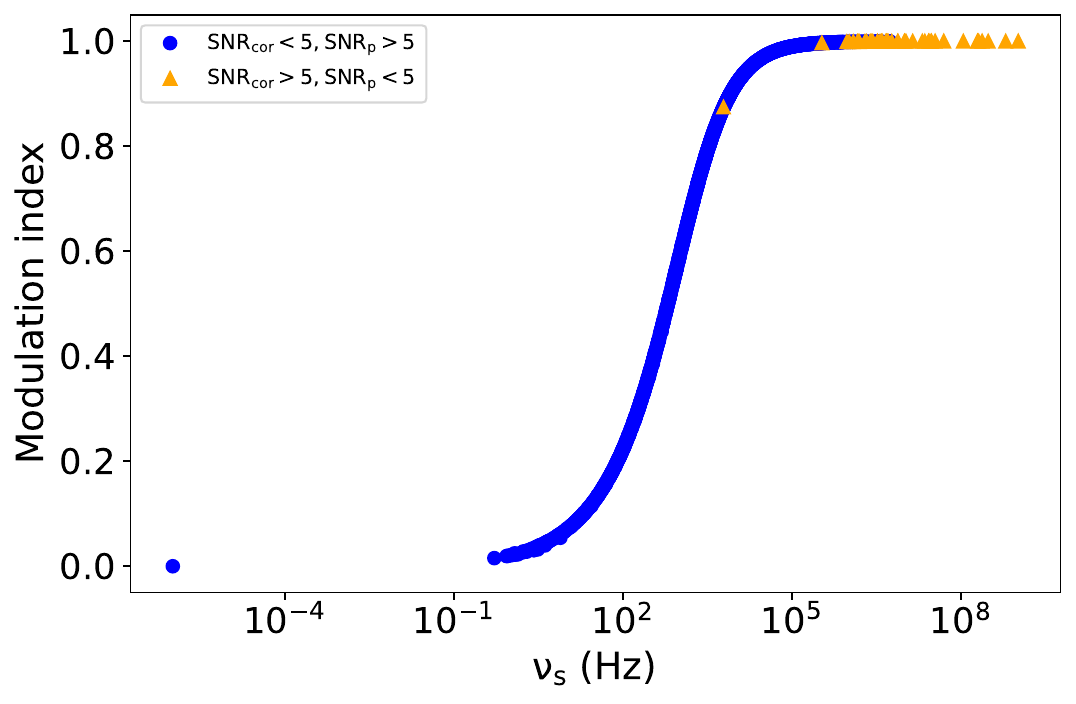}
\includegraphics[width=0.5\textwidth]{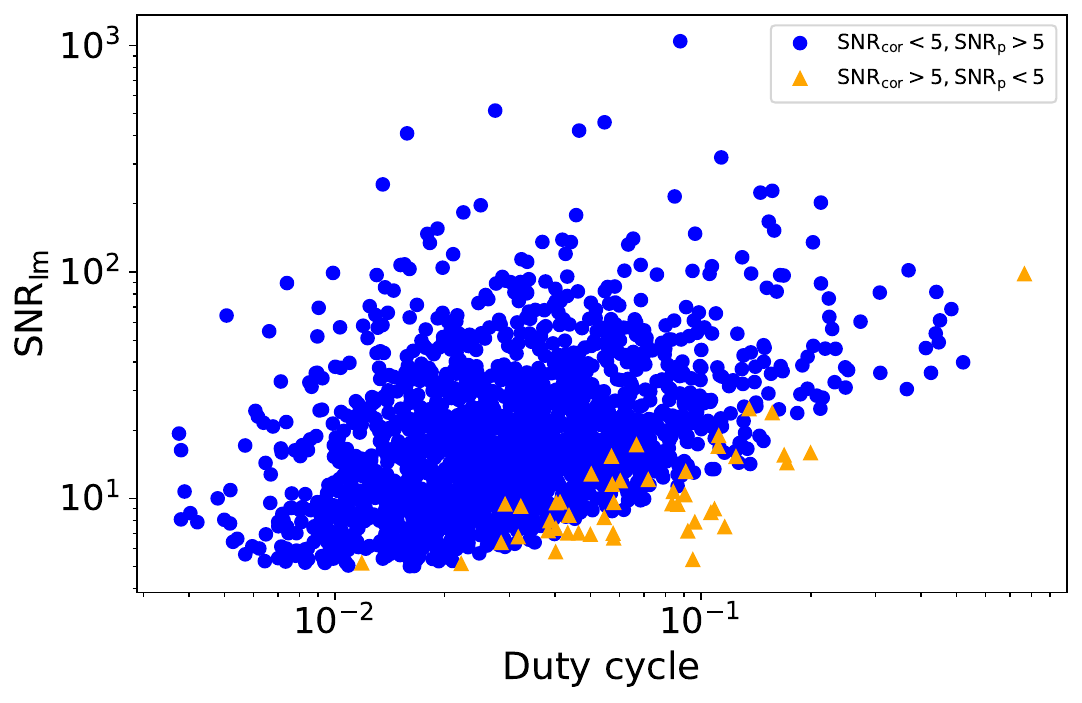}

\caption{\textbf{Left panel}: Modulation index and scintillation bandwidth distributions. \textbf{Right panel}: Relationship between image SNR and duty cycle for normal pulsars detected in time-domain searches but missed in SVCS (blue dots), yellow triangles represent pulsars identified through SVCS but undetected in time-domain searches.}
\label{fig:isnrduty}
\end{figure*}

The normal pulsar population simulation was performed at a central frequency of 610\,MHz, using a bandwidth of 200\,MHz, with a total observation time of 30\,min and a fine temporal resolution of 2\,s. To ensure high spectral sampling, we used 16,384 frequency channels, which allowed high-resolution spectral decomposition of the pulsar signals.\\

Figure~\ref{fig:normpuls} illustrates that certain normal pulsars remain undetected in SVCS yet are identified in time-domain searches. These pulsars generally exhibit characteristics such as a relatively low duty cycle ($<$\,0.09), a low modulation index ($<$\,0.9), a narrow scintillation bandwidth ($<$\,400\,kHz), and a high image SNR ($>$\,10), as shown in Figure~\ref{fig:isnrduty}. In contrast, pulsars detected in SVCS but not in time-domain searches tend to possess a high scintillation bandwidth ($>$\,400\,kHz) and a high modulation index ($>$\,0.9). These sources are typically faint (image SNR $<$ 10) and have a relatively high duty cycle ($>$\,0.09). The right panel of Figure~\ref{fig:isnrduty} shows some pulsars that are exclusively detected in SVCS but are present in the time-domain search region. Although these pulsars demonstrate a theoretical time-domain SNR greater than 9, factors such as degradation and offset lower their overall time-domain SNR below 9. However, their high modulation index ($>$\,0.8) and high scintillation bandwidth ($>$\,40\,kHz) ensure their detectability in SVCS, even as they evade time-domain searches.\\

\begin{figure*}
\includegraphics[width=0.5\textwidth]{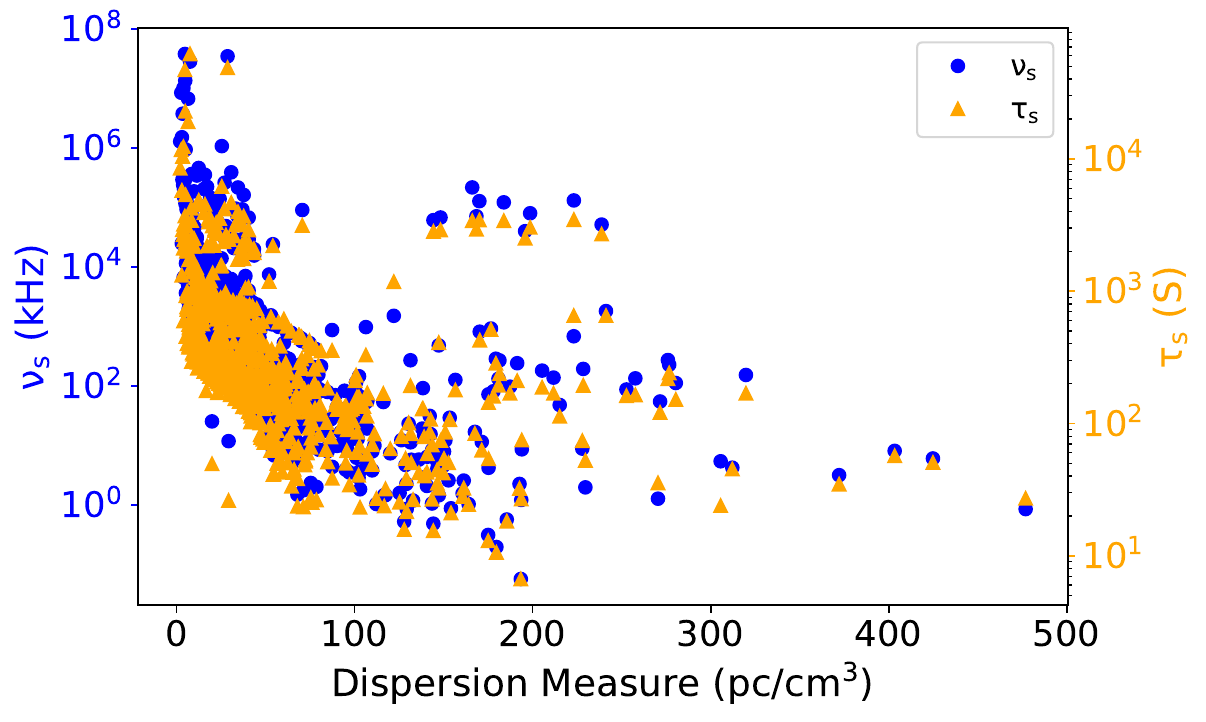}
\includegraphics[width=0.5\textwidth]{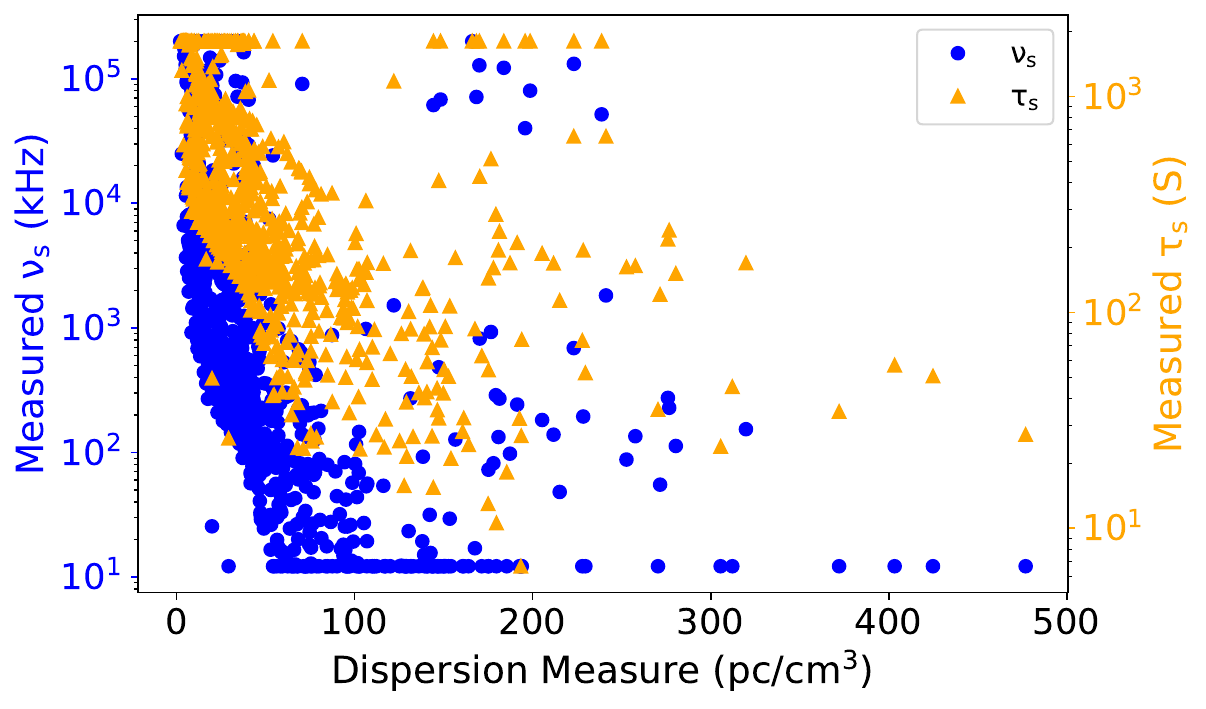}

\caption{The left panel shows the scintillation bandwidth and timescales of all the simulated normal pulsars detected in SVCS as a function of DM. The right panel shows the measured scintillation bandwidth and timescale.
\label{fig:scbw}}
\end{figure*}

In addition, six pulsars are detected through SVCS but remain undetected in the radio image. These pulsars exhibit modulation index $\sim$\,1, scintillation bandwidth $\sim$ the observation bandwidth, and scintillation timescale $\sim$ the integration time. This number may increase, but the upper limits for measurable scintillation bandwidth and timescale are constrained by half of the total observing bandwidth or time. According to Equations~\ref{eqn:image} and ~\ref{eqn:correlation}, even if the image SNR falls below 5 (making them undetectable in the radio image), pulsars can still be detected with SVCS if the scintillation bandwidth or timescale approaches half of the total bandwidth or time, provided that the image SNR exceeds 3. The six pulsars detected here have a DM below 33, aligning with Figure~\ref{fig:scbw}, which shows that low-DM pulsars are associated with higher scintillation bandwidths and timescales.\\

Lastly, pulsars identified in time-domain searches but not in radio images typically have a duty cycle below 0.07. According to Equation~\ref{eqn:pulsar_survey}, when the number of antennas used in the GMRT observation is 26, the SNR$_p$ for pulsars with a duty cycle less than 0.07 surpasses the corresponding SNR$_I$.\\

\textbf{Effect of frequency and time resolution:}\\

\begin{figure}
\includegraphics[width=0.5\textwidth]{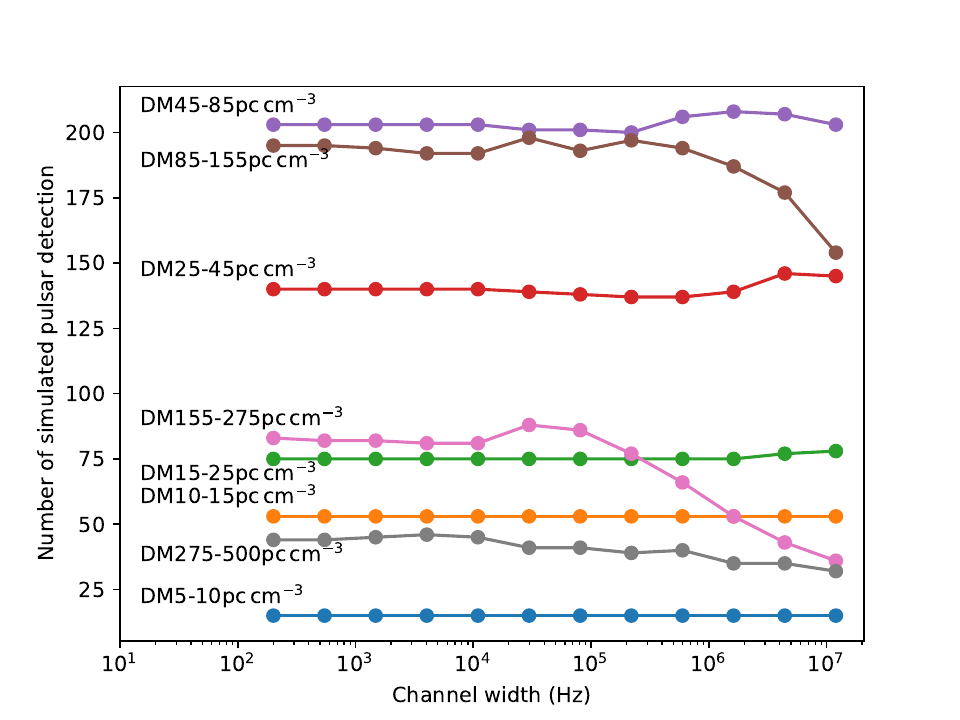}

\caption{The number of simulated normal pulsar detections in SVCS is shown as a function of channel width for different DM ranges. The detection rate decreases with increasing channel width but remains approximately constant up to 10\,kHz, peaks at 20\,kHz, and subsequently declines. These simulations were conducted for a normal pulsar population observed at 600\,MHz with 200\,MHz bandwidth, using 30\,min observations and 2\,s sub-integrations.}
\label{fig:channelwidth1}
\end{figure}

To investigate the effect of channel width on pulsar detection through the SVCS, we simulated pulsar signals across varying channel widths and detected these signals over different DM ranges, as illustrated in Figure~\ref{fig:channelwidth1}. The figure presents the number of simulated pulsar detections as a function of channel width for equidistant logarithmic DM intervals. We find that reducing the channel width increases the detection rate until the channel width reaches approximately 10\,kHz, beyond which further reductions have negligible impact. However, for certain DM ranges, the detection rate decreases slightly as the channel width is reduced. These results show that a channel width around 10\,kHz is optimal to maximize detection efficiency.\\

For the GMRT, the frequency resolution varies from 12\,kHz to 195\,kHz, which typically helps us identify pulsars with a scintillation bandwidth greater than 1\,kHz. 74\% of pulsars have a scintillation bandwidth less than 1\,kHz. This suggests that a higher frequency resolution is needed, but as previous analyses show, increasing the frequency resolution does not result in more detection; on the contrary, the number of detections decreases.\\

For GMRT, the typical time resolution used in imaging is 8\,s and can be as short as 1\,s, which is sufficient to identify pulsars in a diffractive scintillation regime. At 600\,MHz, the full range of the scintillation timescale varies from 0.25\,s to 10\,h, independent of the time resolution and the total duration of observation. 98.2\% of the pulsars detected in SVCS have a scintillation timescale of less than 12\,min at 600\,MHz.\\

\subsection{Millisecond pulsars}
We performed a population synthesis study of MSPs using the \texttt{PsrPopPy} software package, with the High Time Resolution Universe (HTRU) Pulsar Survey \citep{keith2010} serving as our reference dataset. The simulation was normalized to match the detection of 48 MSPs in the intermediate-latitude portion of the southern HTRU survey, resulting in a total Galactic population of 41,000 potentially observable MSPs. This normalization process generated a statistically representative sample of the underlying MSP population in the Milky Way.\\

Our simulation methodology involved adopting the MSP population distribution of the HTRU survey as a fiducial value, subsequently executing a GMRT survey simulation across a diverse parameter space. We systematically varied the observational parameters to explore the detection capabilities and limitations of SVCS.\\

The MSP population simulation was performed at a central frequency of 610\,MHz, using a bandwidth of 200\,MHz, with a total observation time of 30\,minutes and a fine temporal resolution of 2\,s. To ensure high spectral sampling, we used 16,000 frequency channels, enabling high-resolution spectral decomposition of the pulsar signals.\\

\begin{figure}[htbp]
\centering
\includegraphics[width=0.5\textwidth]{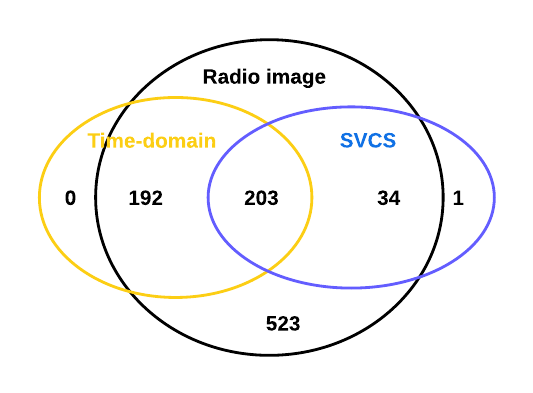}
\caption{Venn diagram for simulated MSPs detected by three techniques. The black, yellow, and blue circles corresponds to MSPs detected in radio images, time-domain searches, and SVCS. Pulsar population simulation is conducted at a frequency of 600\,MHz with a bandwidth of 200\,MHz, an observation time of 30\,minutes and a sub-integration time of 2\,s.}
\label{fig:mspuls}
\end{figure}

\begin{figure*}[htbp]
\includegraphics[width=0.45\textwidth]{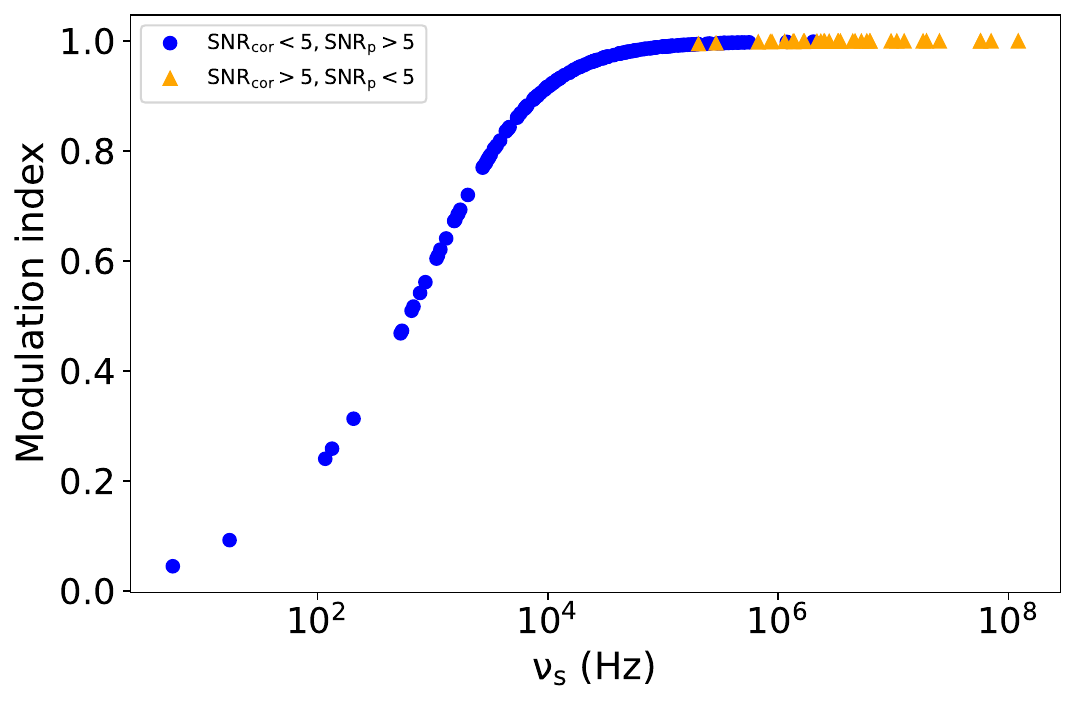}
\includegraphics[width=0.45\textwidth]{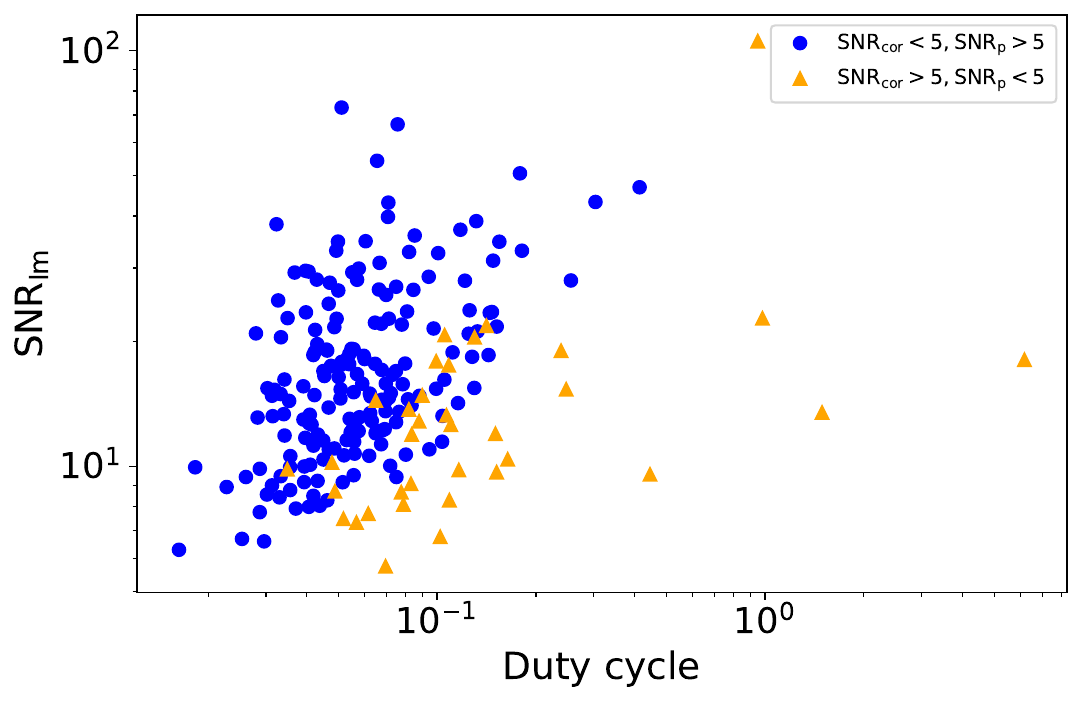}
\caption{The left panel shows the modulation index and scintillation bandwidth of MSPs detected in the time-domain searches but not in SVCS (blue dots). Pulsars detected in SVCS but not in time-domain searches are represented by a yellow triangle. The right panel shows the image SNR and duty cycle of MSPs}
\label{fig:isnrdutyms}
\end{figure*}

Our simulation revealed a significantly higher detection rate for MSPs than for normal pulsars through SVCS. This increased detectability is due to propagation effects in MSPs, particularly a high modulation index and a large scintillation bandwidth, as illustrated in Figure~\ref{fig:isnrdutyms}.\\

The high detection rate of 14\% for MSPs (34 detections out of 238 in SVCS as shown in Figure~\ref{fig:mspuls}) compared to 5\% for normal pulsars (45 detections out of 886 in SVCS as shown in Figure~\ref{fig:normpuls}) can be attributed to multiple factors. Primarily, MSPs have a larger scale height with respect to the Galactic plane, and hence they are typically found at higher Galactic latitudes (and smaller distances), giving them different scintillation properties compared to normal pulsars. They also exhibit distinct spatial and kinematic distributions throughout the Galactic volume compared to normal pulsars. Their concentrated population in the Galactic disk and proximity to the Galactic centre contribute to more favorable scintillation conditions for their detectability through SVCS.\\

\section{Radio continuum Surveys}
\label{sec:radiosurvey}

The next generation of radio telescopes, including the Square Kilometre Array (SKA), the Australian Square Kilometre Array Pathfinder (ASKAP), and the Deep Synoptic Array-2000 (DSA-2000), will produce high-resolution radio astronomical data. Radio surveys using these instruments will map large regions of the sky with a sensitivity exceeding $\sim$ 1\,$\mu$Jy/beam, surpassing previous surveys such as the TIFR GMRT Sky Survey (TGSS), which operated at 150 MHz with a sensitivity of 3.5\,mJy\,beam$^{-1}$ \citep{tgss}. The increased sensitivity and resolution of the new surveys enable the detection of fainter and more compact radio sources, including pulsars that previous surveys could not resolve or distinguish from background sources. Using pulsar population synthesis, we simulate pulsar detection rates in radio images, time-domain searches, and SVCS for these surveys to identify potential new pulsar candidates.\\

\begin{deluxetable*}{lcccccccccc}
\tabletypesize{\scriptsize}
\tablewidth{0pt} 
\tablecaption{We generated simulated populations of 120,000 (41,000) normal pulsars (MSPs) by detecting 1,038 (48) pulsars in Parkes multibeam survey (HTRU survey), then running TGSS, EMU-ASKAP, DSA-2000 and SKA-Mid survey on these generated pulsars for their detection via different techniques. For the simulation, we assumed the sampling time of 0.1\,ms for time-domain searches. The column in the table is survey name, centre frequency, bandwidth, channel width, integration time, sub-integration time, number of antennas, pulsars detected in radio images, time-domain searches, and SVCS.
\label{tab:pulsarsurvey1}}
\tablehead{
\colhead{Survey} & \colhead{Frequency} & \colhead{Bandwidth} & \colhead{Channel} & \colhead{Integration} & \colhead{Sub-} & \colhead{Antennas} & 
\colhead{Radio image} &
\colhead{Time-domain} & 
\colhead{SVCS} & \colhead{Pulsar}\\
\colhead{} & \colhead{} & \colhead{} & \colhead{width} & \colhead{time} & \colhead{integration} & \colhead{} & 
\colhead{} &
\colhead{} & 
\colhead{}  & \colhead{candidates}\\
\colhead{} &\colhead{MHz} & \colhead{MHz} & \colhead{kHz} & \colhead{min} & \colhead{s}
} 
\startdata
TGSS & 147.5& 16.7& 65 & 15& 16.1 & 30& 5114 (152)& 2968 (49)& 523 (41) & 2 (4)\\
EMU-ASKAP & 1300& 300& 1000 & 720& 10 & 36& 4835 (826)& 2724 (227)& 813 (187) & 45 (33)\\
DSA-2000 & 1400& 1300& 134 & 15& 1& 2000& 8403 (2478)& 1181 (130)& 2340 (816) &1306 (688)\\
SKA-Mid & 1400& 810& 50 & 6& 1 & 197& 8293 (1825)& 2483 (256)& 2156 (607) & 519 (366)\\
\enddata
\end{deluxetable*}

\subsection{SKA}
The SKA-Mid is poised to conduct an all-sky continuum survey at 1.4\,GHz, achieving an unprecedented sensitivity of approximately 4\,$\mu$Jy \citep{pradoni2015}. This survey will acquire data over a broad 810 MHz bandwidth with a fine frequency resolution of 50\,kHz. The total observation time is approximately 360\,s, with a time resolution of 1\,s facilitating a detailed analysis. A key strength of this imaging survey lies in the utilization of all 197 SKA-Mid antennas. This antenna deployment allows efficient mapping of the entire field of view, significantly broadening the observable sky compared to time-domain searches.\\

Our simulations demonstrate the efficacy of scintillation-based pulsar detection via SVCS. We simulated 120,000 normal pulsars and 41,000 MSPs. From this simulated population, we identified 8,293 normal pulsars and 1,825 MSPs in the radio images with SNR above 5. SVCS yielded 2,156 normal pulsars and 607 MSPs. Time-domain surveys detected 2,483 normal pulsars and 256 MSPs. We found an overlap of 1,625 normal pulsars and 241 MSPs detected by both scintillation and time-domain survey methods. 12 normal pulsars exhibited scintillation signatures but remained undetected in the radio images themselves. We identified 519 normal pulsars and 366 MSPs solely through SVCS.\\

\subsection{EMU-ASKAP}
ASKAP consists of 36 12-meter antennas spread over a region 6\,km in diameter. The EMU survey of ASKAP will carry out a deep (RMS $\sim$ 10\,$\mu$Jy/beam) radio continuum survey of the entire Southern sky at 1.3\,GHz, covering the entire sky south of $+$30$^o$ declination, with a resolution of 10\,arcsec \citep{norris2011, johnston2008}. The EMU survey has a centre frequency of 1.3\,GHz, a total bandwidth of 300\,MHz, and an integration time of 12\,h for each pointing. The default time and frequency resolutions are 10\,s and 1\,MHz, respectively.\\

From a simulated population of 120,000 normal pulsars, we identified 4,835 using radio imaging techniques. Similarly, we detect 826 MSPs from a simulated population of 41,000. Scintillation alone identifies 813 normal pulsars and 187 MSPs. In comparison, time-domain searches detect 2,724 normal pulsars and 227 MSPs. In particular, there is significant overlap, with 763 normal pulsars and 153 MSPs identified by both methods. Some pulsars, five normal and one MSP, exhibit scintillation detectable in radio images but remain undetected by primary imaging criteria. We identified 45 normal pulsars and 33 MSPs exclusively through SVCS.\\

\subsection{DSA-2000}
The DSA-2000 is a radio interferometer array consisting of 2000 aligned 5-meter dishes distributed across a 19\,km $\times$ 15\,km area in a radio-quiet region of Nevada. Operating in the 0.7-2\,GHz frequency range, the array provides nearly complete uv-plane coverage. Unlike traditional radio telescopes, the DSA-2000 employs a novel imaging system that replaces standard correlator backends with direct imaging capabilities, enabling real-time image formation.\\

The array will conduct a five-year survey covering approximately 31,000 square degrees of sky through sixteen observation epochs. This survey is expected to produce full-Stokes maps with a sensitivity of 500\,nJy/beam, detecting more than 1 billion radio sources. The design and capabilities of the DSA-2000 represent a significant advancement in radio astronomy instrumentation, particularly for large-area sky surveys and time-domain studies.\\

Pulsar population simulations reveal the instrument's detection capabilities. From a simulated population of 120,000 normal pulsars, the DSA-2000 successfully identifies 8,403 pulsars in radio images. Similarly, from 41,000 simulated MSPs, it detects 2,478 MSPs. SVCS uncovers 2,340 normal pulsars and 816 MSPs, while time-domain searches detect 1,181 normal pulsars and 130 MSPs. There is a significant overlap with 1,030 normal pulsars and 128 MSPs detected by both scintillation and pulsar survey methods. The survey simulation revealed 1,306 normal pulsars and 688 MSPs exclusively through SVCS.\\

\subsection{TGSS}
The TGSS is an extensive radio survey between 2010 and 2012. Operating at a central frequency of 147.5\.MHz, the survey observational time is about 2000 hours to map a substantial portion of the sky. The survey coverage extended from $-$55$^o$ declination to the northern polar cap, encompassing an expansive area of 37,000 square degrees through a mosaic of over 5000 partially overlapping pointings. The survey recorded data across 256 frequency channels, spanning a total bandwidth of 16.7\,MHz centred precisely at 147.5\,MHz. Each sky direction underwent an observation, with multiple short snapshots (3 to 5 times) collected during a single night's observation. This approach ensured robust data collection and improved sensitivity. The average integration time per pointing was 15 minutes, with a time resolution of 16.1\,s.\\

The TGSS Advanced Data Release (ADR) achieved exceptional sensitivity and angular resolution, novel for sky surveys at that time. It delivered a median RMS noise level of 3.5\,mJy\,beam$^{-1}$, with angular resolution varying between different declinations. For regions north of declination $>$ 19$^o$, the survey maintained a consistent angular resolution of 25\,arcsec, while adapting the resolution for more southern regions by scaling with the cosine of the declination offset.\\

From a simulated population of 120,000 normal pulsars, the survey identified 5,114 normal pulsars in radio images, including hard-to-detect pulsars such as high-duty-cycle pulsars, binary pulsars, and nearly aligned pulsars. Similarly, of the 41,000 simulated MSPs, 152 were detected. SVCS identified 523 normal pulsars and 41 MSPs, while time-domain searches detected 2,968 normal pulsars and 49 MSPs. 521 normal pulsars and 37 MSPs were detected by both SVCS and time-domain searches.\\

\section{Analysis}
\label{sec:analysis}

\begin{figure*}
\includegraphics[width=0.5\textwidth]{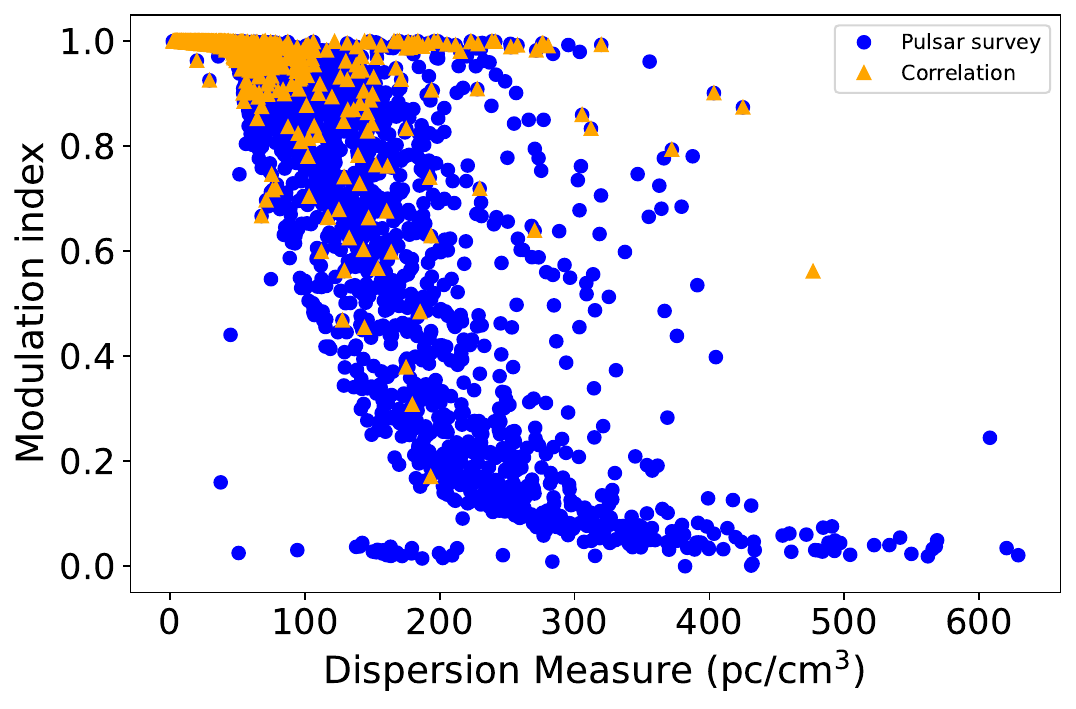}
\includegraphics[width=0.5\textwidth]{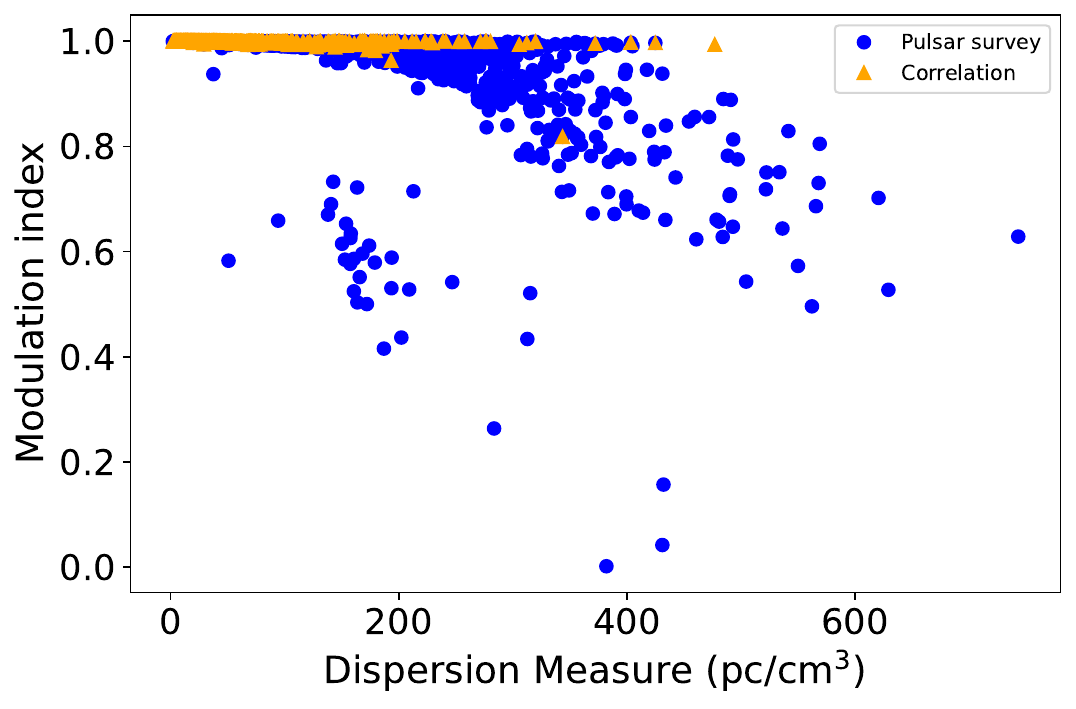}

\caption{Modulation index dependence of DM under different channel widths. \textbf{Left Panel}: Modulation index variation at 12\,kHz channel width. \textbf{Right Panel}: Modulation index variation at 12\,Hz channel width. Color-coded markers delineate detection methodologies: blue dots represent pulsars identified through time-domain searches, while orange triangles indicate pulsars detected via SVCS.
\label{fig:midmhldf}}
\end{figure*}

Figure~\ref{fig:midmhldf} shows the variation of the modulation index with DM, the modulation index exhibits a decline with increasing DM, approaching near-zero values in high DM ranges. This trend directly impacts $\mathrm{SNR_{cor}}$, which scales quadratically with the modulation index, thereby reducing the detection capabilities. The left panel represents a simulation conducted at a channel width of 12\,kHz, a central frequency of 610\,MHz, and a bandwidth of 200\,MHz, revealing that most pulsars exhibit low modulation indices, making them challenging to detect through SVCS. The predominant detection mechanism shifts towards time-domain pulsar searches. In contrast, the right panel uses an extremely narrow 12\,Hz channel width. The modulation indices of most pulsars approach unity, even at high DM.\\

\begin{figure*}
\includegraphics[width=0.5\textwidth]{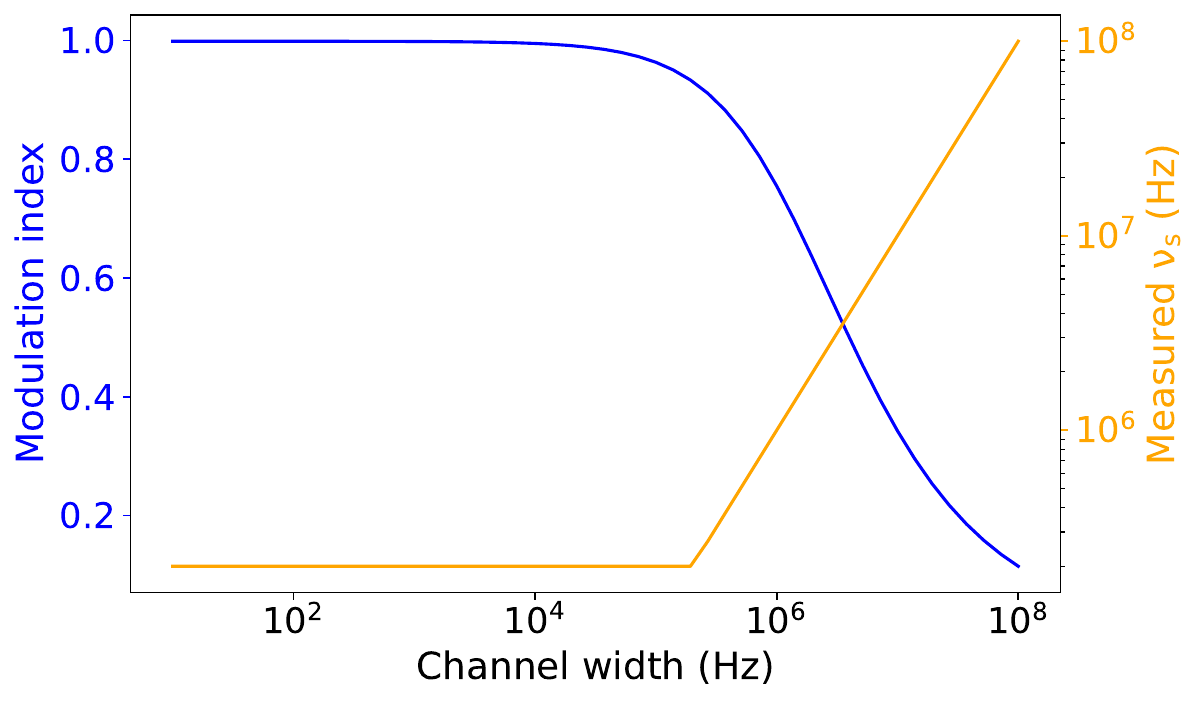}
\includegraphics[width=0.5\textwidth]{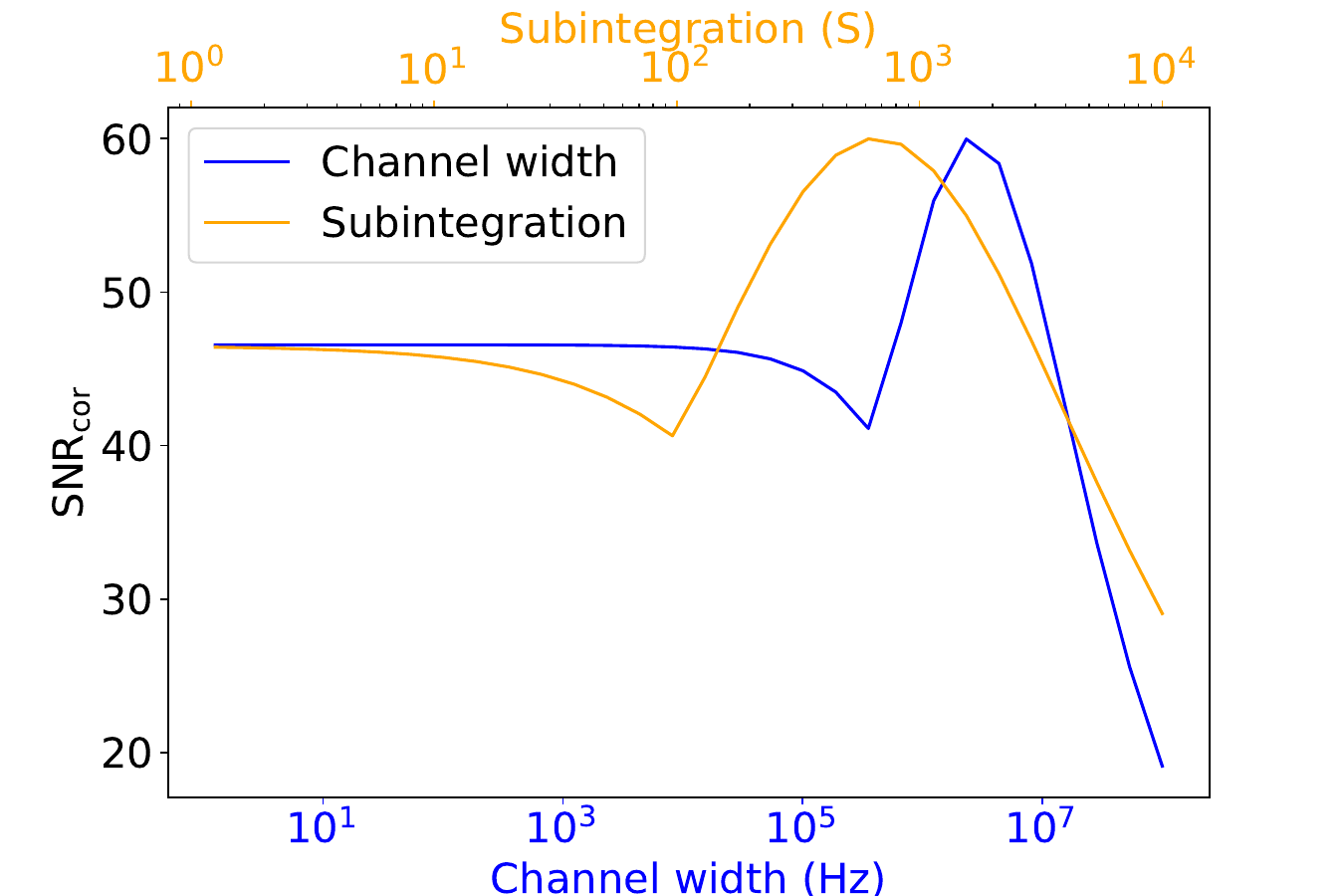}
\caption{\textbf{Left Panel}: Dual-axis representation showing modulation index and scintillation bandwidth variations as functions of channel width. \textbf{Right Panel}: $\mathrm{SNR_{cor}}$ dependence on channel width and sub-integration. Simulation parameters: pulsar scintillation bandwidth $\nu_s$ = 400\,kHz, scintillation timescale $\tau_s$ = 100\,s, channel width = 12\,kHz, sub-integration time = 2\,s, flux = 50\,mJy, rms = 1\,mJy, observation time=30\,min and bandwidth=200\,MHz. 
\label{fig:snrchwd}}
\end{figure*}

The left panel of Figure~\ref{fig:snrchwd} shows that when the channel width is substantially smaller than the intrinsic $\nu_s$, the measured $\nu_s$ equals the intrinsic $\nu_s$ and the modulation index maintains a near-unity value, indicating optimal detection conditions. This regime represents the highest sensitivity for scintillation-based pulsar identification. As the channel width approaches the intrinsic $\nu_s$, a transition occurs: the modulation index begins a decline. This decline becomes particularly pronounced when measured $\nu_s$ becomes equal to channel width, as channel width increases beyond intrinsic $\nu_s$, signifying a detection threshold. For channel widths significantly exceeding intrinsic $\nu_s$, the modulation index asymptotically approaches zero, effectively suppressing scintillation signatures.\\

The right panel of Figure~\ref{fig:snrchwd} illustrates the relationship between the channel width and $\mathrm{SNR_{cor}}$. For channel width significantly smaller than the intrinsic $\nu_s$, $\mathrm{SNR_{cor}}$ maintains a stable value, reflecting optimal detection conditions. This stability arises because the narrow channels fully resolve the scintillation structure. As the channel width approaches intrinsic $\nu_s$, we observe a decrease in $\mathrm{SNR_{cor}}$, indicating a reduced detection sensitivity. This decline reaches a local minimum when the channel width is exactly equal to $\nu_s$, representing a transition point in the detection regime. Beyond this minimum, $\mathrm{SNR_{cor}}$ increases with channel width until it reaches a maximum at approximately 6.67\,$\nu_s$. At this optimal point, the modulation index reaches a value of 0.7, representing an ideal balance between bandwidth averaging and modulation index decline. The specific location of this maximum depends on the parameter $\kappa$ in Equation~\ref{eqn:kappa}. Under these optimal conditions, $\mathrm{SNR_{cor}}$ achieves an enhancement of 48\% compared to its local minimum value, significantly improving the pulsar detection capabilities. An analogous relationship exists with the sub-integration time, where similar considerations apply relative to the scintillation timescale.\\

For the GMRT, the sub-integration time for radio interferometric images spans 2-32\,s, well matched to typical pulsar scintillation timescales, which range from minutes down to approximately 1\,s. This temporal sampling ensures robust $\mathrm{SNR_{cor}}$ measurements.\\

Figure~\ref{fig:channelwidth1} shows that the detection rate remains nearly constant at small channel widths. As the channel width increases, the detection rate improves for certain DM ranges before decreasing sharply at larger channel widths, particularly for DM values between 85-155\,pc\,cm$^{-3}$ and 155-275\,pc\,cm$^{-3}$. This behavior, where the detection rate initially improves with decreasing channel width to 10\,kHz before stabilizing or slightly decreasing below this threshold, can be understood by examining Figure~\ref{fig:snrchwd}.\\

The right panels of Figures~\ref{fig:normpuls} and~\ref{fig:isnrdutyms} display scatter plots of image SNR versus duty cycle. These plots reveal that pulsars detected through SVCS typically exhibit lower image SNR and higher duty cycles compared to those found via time-domain searches. The mean duty cycles are 0.03 for normal pulsars and 0.05 for MSPs in time-domain searches, while SVCS yields higher mean duty cycles of 0.09 and 0.10 for normal pulsars and MSPs, respectively. In the case of MSPs, SVCS detected a few MSPs with duty cycles exceeding 1, demonstrating their effectiveness in identifying pulsars with duty cycles greater than 0.09 where time-domain searches struggle.\\

Table~\ref{tab:pulsarsurvey1} compares the simulated pulsar detection rates in four radio continuum surveys: TGSS, EMU ASKAP, DSA-2000, and SKA-Mid. These surveys were selected to demonstrate the contrast between current and future capabilities in pulsar detection through radio imaging, time-domain searches, and SVCS. The inclusion of TGSS reflects our focus on analyzing GMRT archival data for scintillating pulsars in interferometric observations. The table presents detection rates for three distinct methods: radio imaging, periodic searches, and SVCS. Among these surveys, TGSS achieves the highest normal pulsar detection rate via time-domain searches. This high sensitivity stems from TGSS's operating frequency of 147.5\,MHz, where pulsars' characteristic steep-spectral indices lead to increased flux densities and consequently greater detectability in time-domain searches.\\

The DSA-2000 survey produces the highest detection rates for both normal pulsars and MSPs in the SVCS, exceeding results from other surveys and its own periodic search performance. This outcome arises from the frequency dependence of the scintillation parameters. Scintillation bandwidth and timescale scale with observing frequency following the relations in Equations~\ref{eqn:stime} and~\ref{eqn:sfreq}. The DSA-2000 survey combines a large bandwidth of 1300\,MHz, high sensitivity, and fine time-frequency resolution, increasing the detectability of scintillation-induced intensity variations in dynamic spectra.\\

The SKA-Mid survey achieves high detection rates across all methods, although its SVCS performance lags slightly behind DSA-2000. EMU ASKAP shows intermediate results, with periodic searches favoring normal pulsars and MSPs, whereas SVCS shows lower sensitivity to normal pulsars and MSPs. Radio imaging yields higher detections overall, as it primarily identifies bright, unresolved sources rather than pulsed emission or scintillation signatures.\\

Table~\ref{tab:pulsarsurvey1} shows the simulated pulsar detection rates in radio imaging, periodic searches, and SVCS for four radio continuum surveys: TGSS, EMU ASKAP, DSA-2000, and SKA-Mid. The TGSS survey detects a greater number of normal pulsars through periodic searches because of the steep-spectral indices of pulsars. The steep spectral index increases the brightness of the pulsar at lower observing frequencies, with TGSS operating at 147.5\,MHz. The DSA-2000 survey achieves the highest detection rates for both normal pulsars and MSPs in SVCS among all surveys and also relative to its own periodic search results. This is due to the frequency dependence of the scintillation parameters: scintillation bandwidth and timescale scale with observing frequency as described by Equations~\ref{eqn:stime} and~\ref{eqn:sfreq}. The DSA-2000 survey has a large bandwidth of 1300\,MHz, high sensitivity, and fine time and frequency resolution. These characteristics collectively increase the detection of scintillation-induced intensity variations, which SVCS effectively captures.\\

The SVCS technique can detect at least 24\% of normal pulsar candidates and 50\% of MSP candidates in SKA-Mid observations. For DSA-2000, these fractions increase to 56\% for normal pulsar candidates and 84\% for MSP candidates, while EMU-ASKAP yields 6\% and 18\%, and TGSS detects 0.4\% and 1\% of normal pulsar and MSP candidates, respectively. These pulsars are particularly challenging to identify through time-domain surveys. Some may be nearly-aligned rotators that escape detection in periodicity searches. Among pulsars detectable by both methods in simulations, a subset may be missed in time-domain searches because of the marginal SNR near the detection threshold.\\

\section{Conclusion}
\label{sec:conclusion}
Searching for pulsars in time series is the foolproof method of finding and confirming pulsars. However, some extreme pulsars, such as sub-millisecond pulsars and pulsars in very compact, highly accelerated binary orbits are very difficult to detect via this method. Here, we need to develop some methods that are efficient enough to mitigate the need for large computing resources and fast enough to be applied in real-time for future radio surveys. Searching for pulsar candidates in radio images is one such possibility, as we can identify pulsar candidates on the basis of their steep-spectrum, circular polarization fraction, rotation measure, and time variability.\\

We explored the time-variable nature of pulsars by measuring their scintillation properties and identifying them via SVCS. Through simulations, we have established the operational boundaries of the method in terms of DM ranges and duty cycle constraints, while quantifying its detection efficiency relative to time-domain searches.\\

The primary limitation stems from the inverse relationship between the scintillation bandwidth and the DM, which fundamentally restricts the sensitivity of SVCS at higher dispersion measures. This physical constraint prevents the detection of high-DM pulsars through scintillation signatures. However, the technique provides complementary advantages by identifying pulsar populations that often evade detection in time-domain searches, such as binary systems where orbital motion complicates periodicity analysis, MSPs affected by rapid spin-down, high-duty-cycle pulsars whose broad profiles reduce SNR in folded data, and nearly aligned pulsars whose weak pulse modulation makes time-domain searches ineffective.\\

For such SVCS‑detected candidate pulsars that cannot be confirmed via time‑domain searches, we rely on properties that are uniquely characteristic of pulsars and that effectively rule out other compact radio sources that might mimic variability. Pulsars are the only class of sources known to exhibit DISS. They are compact, steep‑spectrum, and highly polarized radio sources and typically show large rotation measures (RM). By compiling sources that display DISS and further requiring steep spectra, high polarization, and large RM values, we can identify such candidates as pulsars with high confidence, even in the absence of direct pulse detections.\\

\bibliography{sample631}{}
\bibliographystyle{aasjournal}

\end{document}